\documentclass [prb,twocolumn,showpacs,preprintnumbers,amsmath,amssymb]{revtex4}

\usepackage{graphicx}
\usepackage{dcolumn}

\begin{document}
\def\beq{\begin{equation}}
\def\eeq{\end{equation}}
\def\be{\begin{equation}}
\def\ee{\end{equation}}

\def\iomn{i\omega_n}
\def\iom#1{i\omega_{#1}}
\def\c#1#2#3{#1_{#2 #3}}
\def\cdag#1#2#3{#1_{#2 #3}^{+}}
\def\epsk{\epsilon_{{\bf k}}}
\def\Ga{\Gamma_{\alpha}}
\def\Seff{S_{eff}}
\def\dinf{$d\rightarrow\infty\,$}
\def\T{\mbox{Tr}}
\def\t{\mbox{tr}}
\def\cG0{{\cal G}_0}
\def\cS{{\cal S}}
\def\divnum{\frac{1}{N_s}}
\def\vac{|\mbox{vac}\rangle}
\def\intR{\int_{-\infty}^{+\infty}}
\def\intb{\int_{0}^{\beta}}
\def\spinup{\uparrow}
\def\spindown{\downarrow}
\def\bra{\langle}
\def\ket{\rangle}

\def\ka{{\bf k}}
\def\vk{{\bf k}}
\def\vq{{\bf q}}
\def\vQ{{\bf Q}}
\def\vr{{\bf r}}
\def\q{{\bf q}}
\def\R{{\bf R}}
\def\kp{\bbox{k'}}
\def\a{\alpha}
\def\b{\beta}
\def\d{\delta}
\def\D{\Delta}
\def\e{\varepsilon}
\def\ed{\epsilon_d}
\def\ef{\epsilon_f}
\def\g{\gamma}
\def\G{\Gamma}
\def\l{\lambda}
\def\L{\Lambda}
\def\o{\omega}
\def\ph{\varphi}
\def\s{\sigma}
\def\chib{\overline{\chi}}
\def\et{\widetilde{\epsilon}}
\def\hn{\hat{n}}
\def\hnu{\hat{n}_\uparrow}
\def\hnd{\hat{n}_\downarrow}

\def\hc{\mbox{h.c}}
\def\Im{\mbox{Im}}

\def\est{\varepsilon_F^*}
\def\v2o3{V$_2$O$_3$}
\def\uc2{$U_{c2}$}
\def\uc1{$U_{c1}$}


\def\bea{\begin{eqnarray}}
\def\eea{\end{eqnarray}}
\def\#{\!\!}
\def\@{\!\!\!\!}

\def\vi{{\bf i}}
\def\vj{{\bf j}}

\def\+{\dagger}


\def\up{\spinup}
\def\down{\spindown}


\title{Solving Dynamical Mean-Field Theory at very low temperature\\
using Lanczos Exact Diagonalization}
\author{Massimo Capone}
\affiliation{SMC, CNR-INFM, and Dipartimento di Fisica
Universit\'a di Roma ``La Sapienza'' piazzale Aldo Moro 5, I-00185 Roma, Italy,
 and ISC-CNR, Via dei Taurini 19, I-00185 Roma, Italy}
\author{Luca de' Medici\footnote{Present Address: Department of Physics and Center for Materials Theory, Rutgers University, Piscataway, New Jersey 08854, USA}}
\affiliation{Centre de Physique Th\'eorique, \'Ecole Polytechnique,
91128 Palaiseau Cedex, France}
\author{Antoine Georges}
\affiliation{Centre de Physique Th\'eorique, \'Ecole Polytechnique,
91128 Palaiseau Cedex, France}

\begin{abstract}
We present an efficient method to solve the impurity Hamiltonians
involved in Dynamical Mean-Field Theory at low but finite temperature,
based on the extension of the Lanczos algorithm from ground state properties
alone to excited states. We test the approach on the prototypical Hubbard
model and find extremely accurate results from $T=0$ up to relatively
high temperatures, up to the scale of the critical temperature for the
Mott transition. The algorithm substantially decreases the computational
effort involved in finite temperature calculations.
\end{abstract}

\maketitle

\section{Introduction}

Strongly correlated electron systems have received a great deal of attention
in the last twenty years, owing to the interest in classes of materials
such as high temperature superconductors or heavy fermions. Some of the
striking properties of these materials such as strong mass renormalizations,
Mott insulating phases or unconventional magnetic properties are clearly due
to the correlation between electrons, an aspect ignored or poorly taken into
account in conventional band theories. This has led to the development of an
entirely new field and of new theoretical schemes and techniques.
Among those, Dynamical Mean- Field Theory (DMFT)~\cite{revdmft} has emerged as
one of the most powerful, both for model Hamiltonians and as a way to take correlations into account
in realistic electronic structure calculations~\cite{dmftmaterials}.

Within DMFT, spatial correlations are frozen, while local quantum dynamics is fully preserved,
as it happens in the infinite coordination limit, where DMFT becomes indeed the exact
theory.
Under this approximation, a lattice model finds an effective description in terms of 
an impurity model in which an interacting site hybridizes with an effective bath of 
free electrons. The mapping onto the impurity model is enforced by a self-consistency 
condition~\cite{gk_dmft} which contains the information
about the original lattice. The self-consistency equation, as we will see,
connects the hybridization function of the impurity model to the local Green's 
function. 
Therefore we can solve a lattice model within DMFT once
we are provided with a method to solve the impurity model and compute the Green's
function. The Anderson impurity model (AIM), albeit much easier to solve than the original lattice model,
is still a non trivial many-body problem whose solution requires either approximations 
or the use of numerical methods.
Both ``exact'' numerical methods
(Exact Diagonalization (ED)~\cite{ed}, Quantum Monte Carlo (QMC)~\cite{qmc} and the
recently introduced continuous-time version~\cite{ctqmc},
Numerical Renormalization Group (NRG)~\cite{nrg},...)
and approximate ``analytical'' methods
(Iterated Perturbation Theory~\cite{gk_dmft}, the Non-Crossing Approximation~\cite{nca}
and its slave-rotors extensions~\cite{rotors}, the self-energy functional method~\cite{sef}, 
and others) have been successfully employed.
However, most methods have limitations confining them
to either a specific regime (e.g., high temperatures), or to the investigation
of specific physical aspects (e.g., low energy quantities).
In particular, focusing on numerical methods, Hirsch-Fye QMC is well suited
for relatively high temperatures (and weak to intermediate correlations), while
ED based on the Lanczos method has been up to now used only for $T=0$.
The NRG, which uses Wilson's scheme to
solve the AIM ,\cite{nrg} is perfectly suited for an extremely 
accurate determination of the low-energy part of the spectra at zero temperature, but it is 
slightly less accurate on the high-energy part of the spectrum, 
and for finite temperatures.
There is no established reliable tool to deal with the regime of finite but
very low temperature, which is particularly relevant in correlated systems
in which very small energy scales arise, leading to subtle effects (such as
spectral weight transfers) when the temperature is turned on. The aim of this
paper is to introduce a simple modification of the Lanczos strategy in order
to treat the low-temperature regime accurately and with a reasonable 
numerical effort. We emphasize that our approach is different from the
Finite-Temperature Lanczos method developed by Jaklic and Prelovsek\cite{prelovsek,aichhorn}, which is 
built as a tool to use ED at any temperature, but it is in principle exact
only at $T=0$ and in the large temperature limit.
Our method is instead designed to treat the very low-temperature regime with the same accuracy of $T=0$, while it can not be pushed beyond some model-dependent temperature without spoiling the rapidity of the Lanczos algorithm and thus without almost recovering the computational heaviness of the full diagonalization of the Hamiltonian.

As we anticipated in the introduction,
the DMFT method maps a lattice model onto an effective impurity
model that we can write as
\be\label{Ham}
H_{\rm{AIM}}=\sum_{l\s} \e_{l\s} a^\+_{l\s} a_{l\s} + \sum_{l\s} V_{l\s}
(f^\+_\s a_{l\s} + a^\+_{l\s}f_\s) + H_{\rm{at}}
\ee
In this expression, $f^\+_\s$ and $a^\+_{l\s}$ are creation operators for fermions
in with spin $\s$ associated with the impurity site and with the state $l$ of the effective bath,
respectively. For simplicity we consider a single band model, but the formalism
is easily extended to multiorbital models. 
$H_{\rm{at}}$ is the on-site (atomic) part of the original lattice Hamiltonian, which 
contains the interaction terms.
For the Hubbard model,
$H_{\rm {at}}=\varepsilon_f(n^f_\spinup+n^f_\spindown)+U n^f_\spinup n^f_\spindown$, and
(\ref{Ham}) is an Anderson impurity model (AIM).
A fundamental quantity is the so called dynamical Weiss field ${\cal{G}}^{-1}_0(\omega)$, 
which describes the non-interacting part of the effective model, and it
 is related to the Anderson parameters $V_{l\s}$ and $\e_{l\s}$ by the relation
\begin{equation}
{\cal{G}}^{-1}_0(\iomn) = \iomn + \mu - \sum^{N_s}_{l=1} \frac{\vert V_l\vert ^2}{\iomn-\e_l}.
\end{equation}
Introducing the impurity Green's function $G(\tau) = -\langle T_{\tau}c(\tau)c^{\dagger}(0)\rangle$, and its
imaginary-frequency Fourier transform $G(\iomn)$, we can extract the impurity 
self-energy 
\begin{equation}
\Sigma(\iomn) = {\cal{G}}^{-1}_0(\iomn) - G^{-1}(\iomn),
\end{equation}
which within DMFT coincides with the local component of the lattice self-energy.

The self-consistency equation which establishes the equivalence between the lattice and
the impurity models depends on the noninteracting density of states $D(\varepsilon)$ of the original lattice

\begin{equation}
G(\iomn) = \int d\varepsilon \frac{D(\varepsilon)}{\iomn + \mu - \varepsilon - \Sigma(\iomn)}.
\label{selfgeneral}
\end{equation}
 
For an infinite- coordination Bethe lattice
with semicircular density of states of bandwidth $2D$,
(\ref{selfgeneral}) reads:
\be\label{selfcons}
{\cal{G}}^{-1}_0(\iomn) = \iomn + \mu -\frac{D^2}{4}\,G(\iomn).
\ee

A practical solution of DMFT consists of an iterative solution of the
impurity model. Starting from a given choice of the Weiss field, the 
impurity Green's function has to be computed with some ``impurity solver''.
The knowledge of $G$ allows to compute $\Sigma$ from which, exploiting 
the self-consistency condition (\ref{selfgeneral}) one finds a new
Weiss field. The process is then iterated until convergence.

Let us now briefly recall the basic idea behind using ED as an impurity solver
in the DMFT context. ED requires a truncation of the sums over $l$
in Eqs.~(\ref{Ham}) and (\ref{selfcons}) up to a finite value $N_s$,
the exact hybridization function  being recovered in the limit $N_s\rightarrow\infty$.
The accuracy of this method thus relies on how closely one can reproduce an
infinite- $N_s$ bath with a finite- $N_s$ one.
To be concrete, our discretized impurity model reads exactly as Eq. (\ref{Ham}),  
with a small value of $N_l$. We can view this as a finite number of ``sites'',
each directly hybridized with the impurity, in the so-called ``star'' geometry.\cite{notageometrie}
At every iteration, once  ${\cal G}^{-1}_0$ is obtained through the self-consistence
equation, the new set of Anderson's parameter is obtained through a fitting
procedure, where a functional distance between the ${\cal G}^{-1}_0$ coming from
Eq. (\ref{selfcons}) and a discretized version is minimized. In this work
we minimize the function
\begin{equation}
\label{distance}
\chi = \sum_n W(\iomn) \vert {\cal G}_0(\iomn) - {\cal G}^{N_s}_0(\iomn)\vert, 
\end{equation}
where ${\cal G}^{N_s}$ is the inverse of the discretized version of the Weiss field, 
the norm $\vert \ldots \vert$ is the square root of sum of the squares of the differences of the
real and imaginary parts, and $W(\iomn)$ is a weight function. In this work we
take the flat function $W(\iomn)=1$, but more selective functions can be useful
for specific problems. For example, one can give more weight to small frequencies
using $W(\iomn) = 1/\omega_n$.\cite{1d}
The truncation error measured by $\chi$ is the only systematic error in the ED 
solution of DMFT. As shown in Ref. \cite{revdmft}, $\chi$ decreases exponentially
by increasing the number of levels $N_s$, so that relatively small numbers provide
accurate information.
The method is also able to provide real-frequency quantities without the need of
analytical continuation tools, even if the spectra are necessarily discrete, due to
the discreteness of the effective model. Nevertheless, many informations about
single-particle and optical spectra can be obtained, mainly as far as the evolution
of spectral weight is concerned.\cite{toschi}

In order to access finite temperature properties, one needs in principle the full
spectrum of the system. Therefore, the size of the matrix to be diagonalized
($4^{N_s+1} \times 4^{N_s+1}$) poses severe limitations on the values of $N_s$ which
can be handled. Even using all the symmetries of the Hamiltonian, one can hardly go beyond
$N_s=5,6$ using full diagonalization.
According to (\ref{selfcons}), when self-consistency is achieved, the
hybridization function of the bath is proportional to the local Green's
function for a Bethe lattice. Therefore, a rough approximation of the bath for small $N_s$
is equivalent to a poor
description of $G(i\omega)$~\cite{notabethe}. This is expected to be more
relevant at low temperature, where the Green's function is more structured,
while at high temperature some structures can be broadened and eventually
washed out. Notice that, nonetheless, the value of the energies of the 
Anderson impurity model can adjust, and in particular, their value can become
arbitrarily small, as well as the weights can vanish. In this way, the method
is able to describe, e.g., the Mott transition, where the Fermi-liquid
coherence energy goes to zero.
It is therefore desirable to increase $N_s$ in order to obtain a reliable
description of the low-temperature region. A standard way to increase $N_s$
is to replace a full diagonalization with the Lanczos algorithm~\cite{golubvanloan}.
In this method one builds an orthonormal basis in the subspace spanned by
the vectors $\vert \phi\rangle,H\vert \phi\rangle,
H^2\vert \phi\rangle,...,H^{N_l}\vert \phi\rangle$, where $\vert \phi\rangle$
is an arbitrary initial state with non-zero overlap to the
groundstate. One can see that in this basis the Hamiltonian becomes tridiagonal
and that even severely truncating the basis (i.e. limiting the number of states
 in the Lanczos basis $N_l$) the
lowest lying states converge to the exact ones very quickly as a function of
$N_l$. In practice the groundstate is very well converged already for basis
of the order of $N_l\sim 100$ even for huge matrices of size of the order of
millions. Further increasing $N_l$, the low-lying excited states gradually
converge with a speed which basically depends on the energy distance
between those states. More care has to be taken to properly handle 
degenerate states and their multiplicity, as we briefly discuss in Sec. \ref{control}.

Due to these convergence properties, this method has been up to now
used  mainly for the investigation of zero-temperature properties,
for which only the groundstate vector needs to be determined.
Nevertheless, Lanczos diagonalization can in principle still be used
at finite (but low enough) temperatures, for which just a few
low-lying states are needed to describe the system.
In this work, we demonstrate that such an extension
of the Lanczos algorithm can be used successfully in the DMFT context.
The next subsection describes our approach.

\subsection{Extension to Finite Temperature}

As we anticipated previously we present here a rather straightforward
extension of the Lanczos scheme to finite temperature, in which a 
small number of excited states of the Hamiltonian matrix are computed.

We first show that the relevant quantities of the AIM can be
expressed as a sum over the eigenstates of the Hamiltonian, each with a
Boltzmann factor which weights it according to the energy distance from the 
groundstate. Thus the sums can be
truncated to a finite number at low- enough temperatures.
This observation is trivial for the partition function
$Z=\sum_n e^{-\b E_n} \label{part_f}$.
For the impurity Green's function, we start from the usual spectral
representation:
\be
\label{gtrunc}
G_{\sigma}(\iomn)=\frac{1}{Z}\sum_{m,n} \frac{\left\vert \langle m\vert f^\+_\s\vert n\rangle\right\vert ^2}
{E_m-E_n-\iomn}[e^{-\b E_n}+e^{-\b E_m}] 
\ee
in which $\vert n\rangle$ and $E_n$ are eigenvectors and eigenvalues of $H_{AIM}$.
This can be recast in the form:
\be\label{G_sum_Gm}
G_{\sigma}(\iomn)=\frac{1}{Z}\sum_m e^{-\b E_m} G_{\sigma}^{(m)}(\iomn)
\ee
where:
\be\label{G_m}
G_{\sigma}^{(m)}(\iomn)\equiv \sum_n
\frac{\left\vert \langle n\vert f_\s\vert m\rangle\right\vert ^2}{E_m-E_n-\iomn} +
\sum_n \frac{\left\vert \langle  n\vert f^{\dagger}_\s\vert m\rangle\right\vert ^2}
{E_n-E_m-\iomn}
\ee
The ``partial'' Green's function $G_{\sigma}^{(m)}(\iomn)$ involves creating (or destroying)
a particle into state $\vert m\rangle$.
It can be readily calculated from $\vert m\rangle$ alone, without 
knowing the whole spectrum spanned by $\vert n\rangle$, just like the $T=0$ 
Green's function (which is simply $G_{\sigma}^{(0)}(\iomn)$) can be computed 
from the ground-state~\cite{notalanczos}.
The exponential factor in Eq.~(\ref{G_sum_Gm}) indicates that at large enough
$\b$ only a small number of eigenstates needs to be calculated, and the sums
can be limited up to $n = N_{kept}$.
Every other spectral quantity, like, e.g., the dynamical spin susceptibility
can be cast in an analogous form exploiting the
Lehmann spectral representation.

The calculation of a few excited states (and hence investigating very low
temperatures) is obviously possible, even if it is not as straightforward
as the evaluation of the pure groundstate.
What is far less obvious is whether a reasonably manageable number
of states is enough to access the temperature range in which the
physical properties of the system start to deviate significantly from $T=0$
(e.g., reaching the Fermi liquid coherence scale in the correlated metal).
The answer to this question depends on the model and on the range of
parameters, since it is mainly connected
with the level spacing in each subsector with given quantum numbers.
In this work, we address this question using as a benchmark test the
half-filled Hubbard model.
This amounts to iteratively solve the AIM (\ref{Ham}), computing
the Green's function, and determine from it a new set of
parameters $\varepsilon_l,V_l$ by minimizing the difference between the two
members of Eq. (\ref{selfcons}). Then the new AIM is solved and the procedure
is iterated until convergence.

We demonstrate that the finite-T Lanczos procedure
can be applied quite successfully to the investigation of the Mott transition
region at finite temperature. Our main results are that:
{\it (i)} The method allows us to easily solve the model for $N_s=6$ at
a considerably lower computational cost than full diagonalization
{\it (ii)} Within finite-T Lanczos, larger
values of $N_s$, ($N_s=8,9$ and in principle the same values that are accessible at $T=0$) can be used,
which require a huge computational effort using the standard 
 full ED, where the Hamiltonian is fully diagonalized.
{\it (iii)} Using $N_s=8$ we can draw the phase diagram of the Mott transition
up to temperatures close to the Mott transition point at a
reasonable computational cost (i.e., keeping a relatively small number of
states).

We finally briefly comment on the difference between our use at finite temperature
of the standard Lanczos algorithm and the well established finite-temperature Lanczos 
method developed by Jaklic and Prelovsek\cite{prelovsek}. 
The method discussed in Refs. \onlinecite{prelovsek,aichhorn} is an ingenious modification
of the Lanczos algorithm, in which the thermal averages are obtained as averages
over random samples of shortened Lanczos chains. The original version of the method\cite{prelovsek}  provides
remarkably good results in the relatively high-temperature regime, but it is not 
particularly efficient at low temperatures, and modifications have been proposed
to overcome this limitation\cite{aichhorn}. On the other hand, our method is 
precisely built to provide basically exact results for low temperature, and it
certainly breaks down (or becomes infeasible) at some temperature.
Thus, the two approaches are basically complementary.

\subsection{Control of the Approximation}
\label{control}
Since the main limitation of the Lanczos method comes from memory
requirements, our finite-$T$ implementation can in principle solve matrices
of the same size as for $T=0$. In practice the evaluation of excited
states naturally slows down the method, ultimately limiting the number of 
states we can
handle. Notice that the computation of excited states does not only require
a larger number of Lanczos steps, but it is further plagued by a loss of
orthogonality in the Lanczos basis, which gives rise to the so-called
"ghost states", i.e., to replicas of the converged vectors with small
weight. Different procedures have been devised to overcome this problem,
mainly based on selective reorthogonalization\cite{golubvanloan}.
This necessary complication of the algorithm leads to an increase of
computational time, which depends on many details of the spectrum.

We finally mention a potential limitation of the present approach, which 
descends from the relative capability of the Lanczos method to handle
degenerate states. It is not difficult to realize that, if the matrix we
try to diagonalize has a degenerate spectrum, the algorithm is not 
able to separate the different states, and to properly determine the 
multiplicity. The simplest way to avoid this problem is to implement
all the symmetries of the Hamiltonian, and to diagonalize independently
the Hamiltonian matrix in each symmetry subsector. 
In this case all the degeneracies associated to those symmetries can not plague the calculation, as the degenerate states will appear in separated subsectors. 
Therefore only the errors arising from accidental degeneracies can affect the
accuracy of our calculation. At least in the case we discuss here, we hardly 
encounter any measurable effect of such degeneracies, even if we can not completely
rule out such effects in other models.
\vspace{0.6truecm}
\begin{figure}[htbp]
\begin{center}
\includegraphics[width=7.5cm]{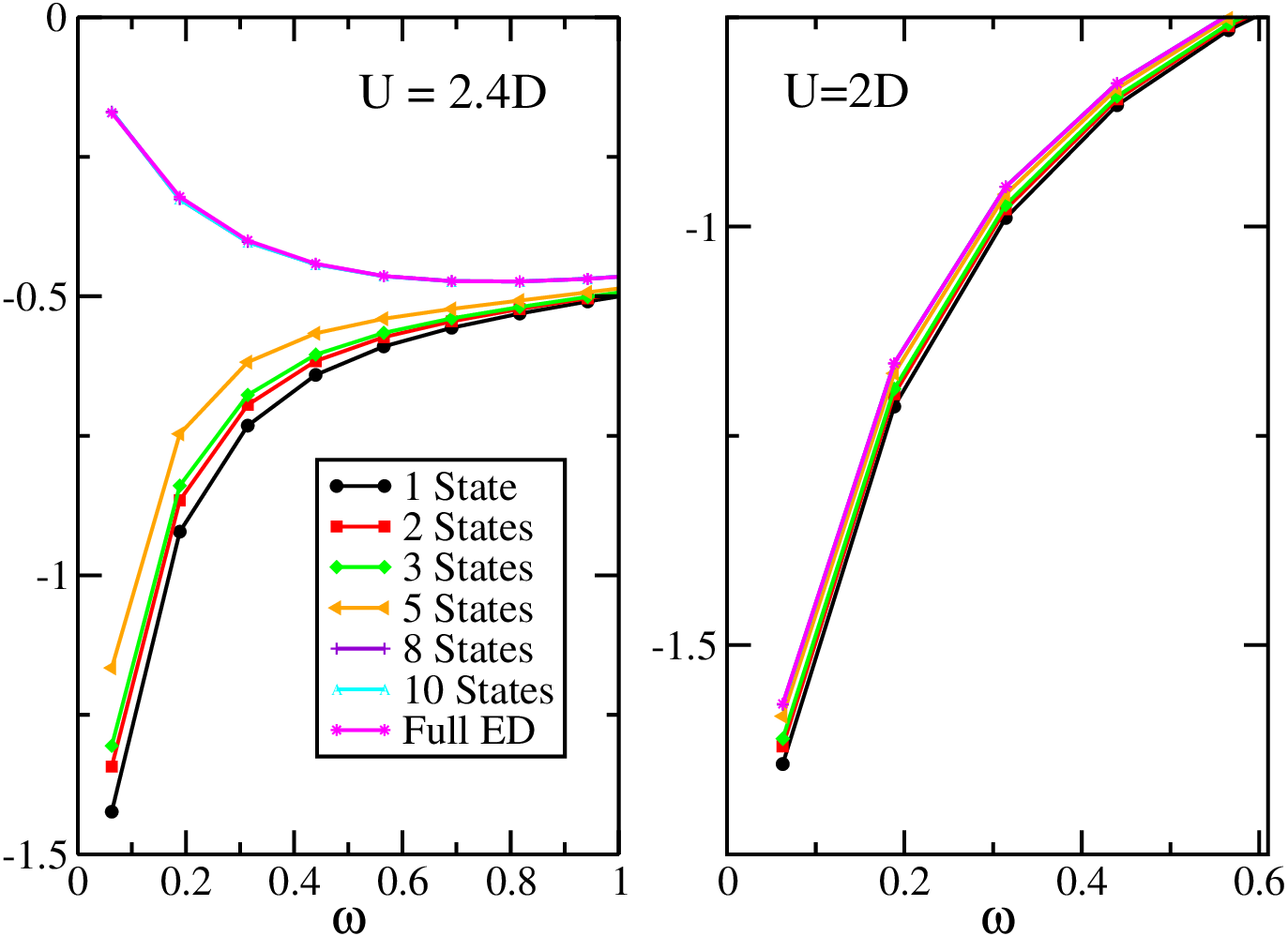}
\end{center}
\caption{(Color online) Imaginary part of the local Green's function on the Matsubara
axis for different values of $N_{kept}$ compared with the full diagonalization
of the Hamiltonian matrix for $N_s = 6$, $\beta=50/D$.
The left panel shows  $U=2.4D$ and the right one  $U=2D$.
}
\label{fig1}
\end{figure}

Our approach has two sources of error, whose effects can be minimized 
in a partially conflicting way. The first is the standard discretization of the
Weiss fields, measured by the value of the distance $\chi$ defined in (\ref{distance}), which is more relevant for low temperature, and the second is
the truncation in Eq. (\ref{gtrunc}), which will obviously be more and more
relevant as $T$ is increased.

\begin{figure}[tbph]
\begin{center}
\includegraphics[width=7cm]{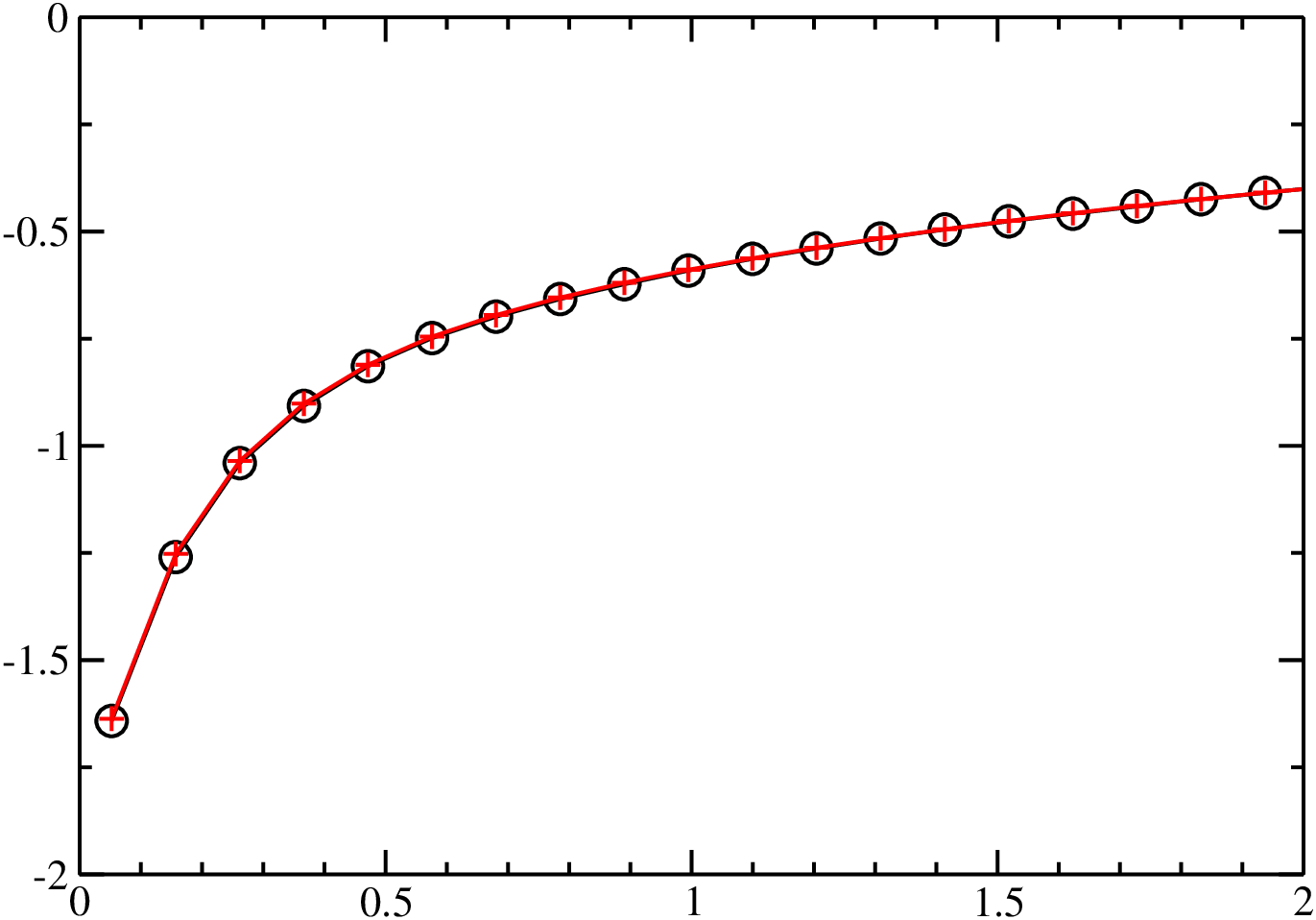}
\end{center}
\caption{(Color online) Imaginary part of the local Green's function on the Matsubara
axis from finite-T ED ($N_s=8$, $N_{kept}=40$) and Hirsch-Fye QMC\cite{jan} for $\beta=60$ and $U=2D$}
\label{fig2}
\end{figure}

We start by discussing the effect of the second kind of truncation,
since the first has been already discussed in the literature, and it has been
shown to be rather benign\cite{revdmft}.
We show results for the paramagnetic half-filled Hubbard model, 
and we compare $G(\iomn)$ (we drop henceforth the spin index)
obtained from full diagonalization of the Hamiltonian matrix for $N_s=6$, which
is basically the maximum size which can be treated with full ED,
with our results for
different values of $N_{kept}$, for a relatively high temperature $T=1/\beta=1/50$.
It is important to underline that $N_{kept}$ is the total number of states in 
the full Hilbert space. Obviously the states will belong to different subsectors
with given quantum numbers. This means that in each subsector the number of 
converged states will be significantly smaller, hence the calculation 
relatively fast.
The comparison, reported  in Fig.~\ref{fig1} clearly shows that
the convergence of our method as a function of $N_{kept}$ is extremely fast,
and the results are indistinguishable from the exact ones
already for $N_{kept}=10-20$.
Lowering the number of kept states, the result approaches the $T=0$ one.
We notice that the results converge fast to the
exact ones irrespective of the value of the interaction, even in a
``difficult'' case such as $U=2.4D$,
for which the $T=0$ solution is a metal while the system is insulating at
$T=1/50$ (convergence is much smoother at e.g $U=2.0 D$).
The inclusion of a few excited states is therefore enough to
qualitatively modify the physics of the system.
It is important to emphasize the significantly lower computational time
of our method, in comparison to the full diagonalization.
In our implementation of the selective re-orthogonalization, we gain a factor
of $\sim 10$ for $N_{kept}=20$ and $\sim 20$ for $N_{kept}= 10$ with respect to full diagonalization (The precise 
numbers depend on many details of the spectrum). In practice, the method 
only introduces a factor of around 3 for $N_{kept}=20$
 in the computational time with respect to $T=0$ ED, so it still substantially
faster than QMC methods.
This benchmark of our approach allows us also to determine a criterion for
stopping the inclusion of excited state. We define a ``difference'' introduced
by the inclusion of the $n^{th}$ state, as $D_n = \sum_{\iomn} \vert G^{n} - G^{n-1}\vert$ ($G^n$ being the Green's function obtained by including $N_kept=n$ states, i.e. $n$ terms in the sum in Eq. (\ref{G_sum_Gm}))
and stop when this distance becomes smaller than a given tolerance, whose value
can be extracted from the comparison with full ED for $N_s \le 6$ and exported to
larger $N_s$ values where the full ED is not feasible.
This is a first indication that perfectly affordable calculations provide
essentially exact results from $T=0$ up to finite temperatures of physical interest.
In particular, we used our method for $N_s=8$, where the size of the Hilbert
space inhibits, or makes extremely heavy, the full diagonalization of the
Hamiltonian.
The results, reported in Fig.~\ref{fig2}, are definitely satisfactory.
The Green's function obtained with our method for
$U=2D$, $\beta=60$ and $N_{kept}=40$ is basically indistinguishable from QMC solution
for the same physical parameters.

\vspace{0.6truecm}

\section{Results}
In order to prove that our algorithm works in a wide range of parameters, we now
draw a phase diagram for $N_s=8$ and $N_{kept}=40$. This relatively
high number of excited state has been chosen according to the criterion discussed previously. More precisely we obtain $D_{40} < 10^{-8}$ for the highest temperature we consider $T=0.02W$, and obviously even smaller values for the lower temperatures.
We notice that a larger number of states has to be used here with
respect to $N_s=6$, due to the larger Hilbert space.
The scenario for the  Mott transition in the paramagnetic sector in
the Hubbard model is now well established. Two distinct solutions
exist, with metallic and insulating character. The former exists for
$U$ smaller than a temperature-dependent value $U_{c2}(T)$, and the
latter for $U > U_{c1}(T)$. At $T=0$ the transition is of second order
and takes place at $U=U_{c2}(0)$, while it becomes of first order
at finite temperature. The coexistence region $U_{c1} < U < U_{c2}$
shrinks as the temperature is increased and closes at a critical
temperature $T_c$, where the first- order line ends in a critical point.
From a practical point of view it turns out easier to determine the numerical value of  $U_{c2}(T)$ line by computing the local spin susceptibility, a quantity which dramatically changes at the transition
point from a Pauli-like susceptibility in the metal to a large ($\propto 1/T$) value associated to local moments in the insulator. This is physically related to the increase of the effective mass when the metallic behavior is lost. The characterization of $U_{c1}(T)$ requires more care. While, like $U_{c2}(T)$, this line is associated to the disappearance of a metastable solution, there is no obvious quantity with a critical behavior when this line is approached. In practice, at each $\beta$, we moved from large to small $U$ with extremely small steps, until the insulating solution disappears.
\vspace{0.6truecm}
\begin{figure}[htbp]
\begin{center}
\includegraphics[width=7cm]{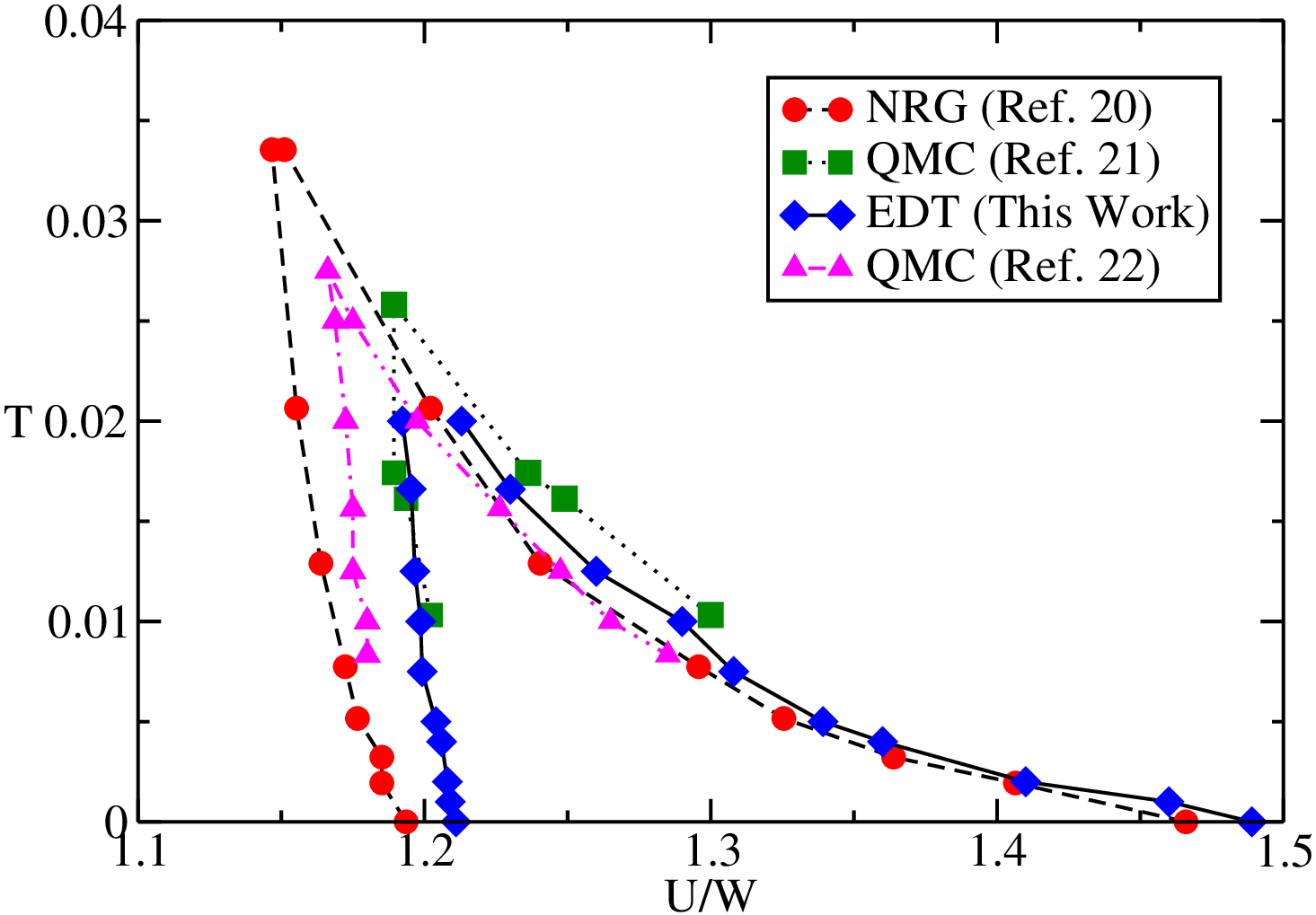}
\end{center}
\caption{(Color online) Phase diagram for the Mott transition in the
paramagnetic sector obtained through finite-temperature ED with $N_s=8$ and
$N_{kept}=40$, compared with previous estimates by Numerical Renormalization
Group of Ref. \cite{bulla} and Quantum Monte Carlo of \cite{udo} and \cite{blumer}.
The ED curves are stopped at the highest temperature where the chosen number
of states determined a negligible truncation error.}
\label{fig3}
\end{figure}
Fig.~\ref{fig3} presents our results for this
phase diagram, and a comparison with the Numerical Renormalization Group
results of Ref. \cite{bulla} and the QMC results of \cite{udo} and \cite{blumer},
which are used as references of popular methods used to study the Mott transition.
We did not add the results of other approaches (self-energy functional, continuous time Monte Carlo, \ldots)
in order to make the figure more readable, and
we emphasize that the aim of this comparison is to prove the ability of our approach
to study finite-temperature properties accurately, rather than a detailed comparison
with different approaches.
The reported data clearly show that our method not only reproduces the
Mott transition scenario at a qualitative level, but also provides results in
extremely good quantitative agreement with established methods.
In particular our method is extremely close to NRG at low temperatures,
where this approach is basically exact, and it is in very good agreement
with QMC at higher temperature, where the latter method becomes accurate.
Our method therefore accurately bridges between the most popular well
established impurity solvers, and allows to span a sizable region of the
phase diagram with good accuracy with a single approach.
We find that $N_{kept}=40$ produces a negligible truncation error up to $\beta=50$,
where our solution is still in extremely good agreement with previous results.
Unfortunately, it is apparently difficult to get closer to the Mott
endpoint, where the number of states needed to get a reasonable accuracy
becomes larger and larger, due to the critical fluctuations which tend
to diverge as the critical point is approached.
In principle one gets a Mott endpoint also with 40 states,
but the large truncation error suggests us not to plot the data around this
point, where the method becomes less reliable, at least quantitatively.

\begin{figure}[htbp]
\begin{center}
\includegraphics[width=7cm]{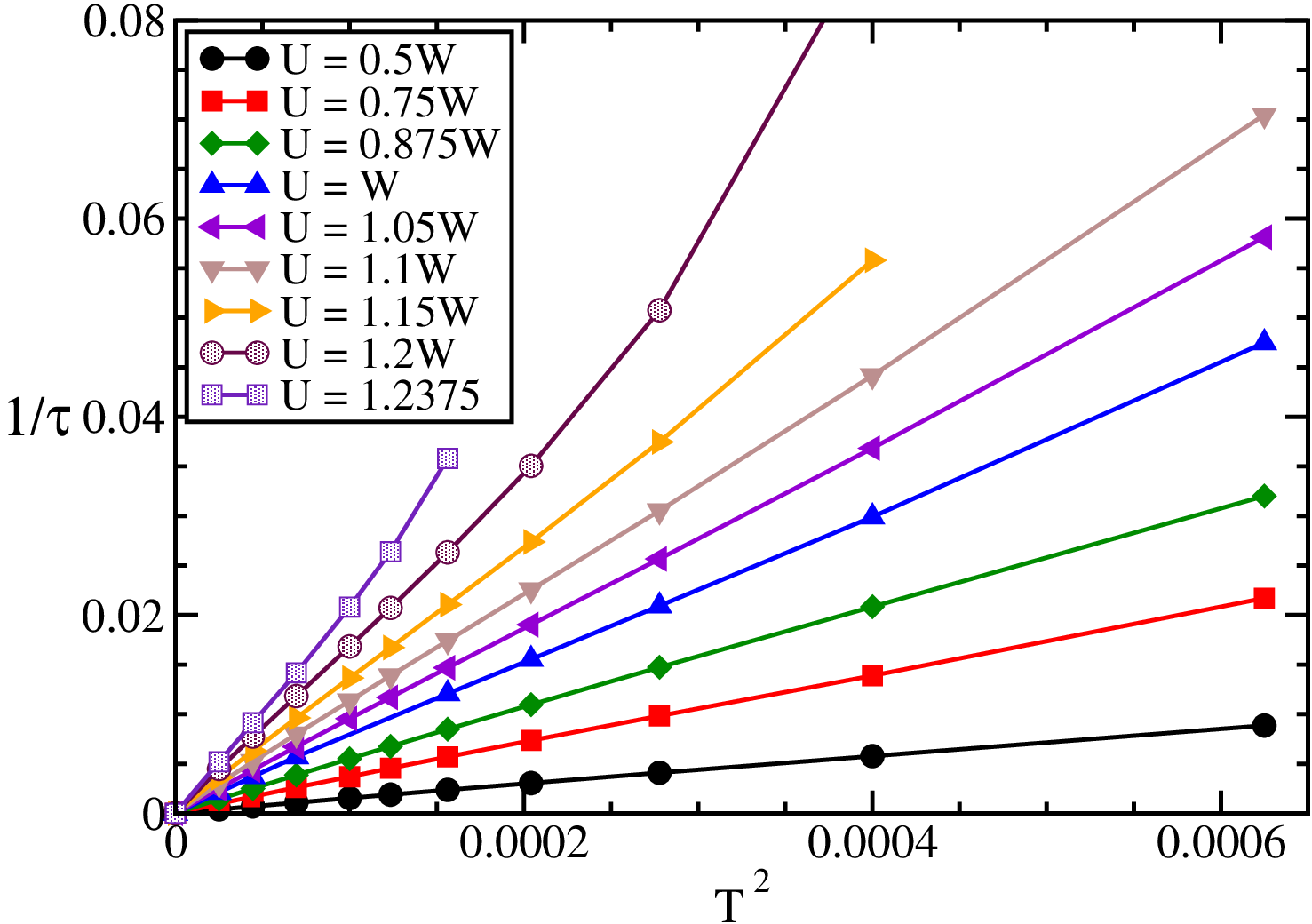}
\end{center}
\caption{(Color online) Inverse quasiparticle lifetime $1/\tau$ for $N_s=7$ as a function of square temperature for the displayed values of the correlation strengths. The linear behavior in $T^2$ characteristic of the Fermi liquid is apparent, with a slope that increases with $U/W$}
\label{fig4}
\end{figure}

A confirmation of the ability of our method to accurately describe the low-temperature regime, we calculated the inverse lifetime of the quasiparticles $1/\tau = Z_{qp} lim_{\omega\to 0} Im\Sigma(i\omega)$, where $Z_{qp} = (1-\partial Re\Sigma(\omega)/\partial\omega)^{-1}$ is the quasiparticle weight.
It has been shown that the metallic phase of the Hubbard model studied in DMFT is a Fermi-liquid. According to Landau Fermi-liquid theory, $1/\tau$ has to be proportional to  $T^2$ at low temperatures, with a coefficient which increases as we approach the Mott transition. Our method correctly reproduces this behavior with limited computational effort, as shown in Fig. \ref{fig4}. This result is not easily accessible to standard methods, and it shows precisely the main virtue of our approach, which works at its best in the low-temperature regime, where Fermi-liquid behavior, and possible violations are directly and unambiguously detectable.
\vspace{0.6truecm}

\section{Conclusions}
We have shown that a finite- temperature extension of the Lanczos algorithm can be successfully applied to the solution of the self-consistent impurity model appearing in DMFT. The inclusion of a few excited states allows for a reliable description of the physics up to temperatures of the order of the Mott critical endpoint with a relatively small increase of computational cost with respect to the well-established $T=0$ exact diagonalization.
The method allows for a computationally cheap investigation of single-band models with extreme accuracy at very low temperatures, where the Fermi-liquid behavior and its possible violations can be investigated unambiguously.
Furthermore, the present approach opens the way to the use of ED as a (small) finite temperature solver for more timely lines of researches, like cluster extensions of the DMFT\cite{cdmft,dca}, or realistic calculations of properties of correlated materials~\cite{dmftmaterials}, at basically the same computational cost of $T=0$ studies. As we already discussed the present algorithm has indeed the same memory requirements as the $T=0$ standard approach

We notice that, despite the scaling of ED methods with the number of orbitals (or, equivalently, sites in the cluster) is not favorable, ED has been successfully applied at $T=0$ to three-orbital models\cite{c60} and to CDMFT for a 2$\times 2$ plaquette\cite{plaquette}. These implementations require a total number of levels of the order of $N_s \simeq 12$. As we discussed in Sec. \ref{control}, the present approach can be applied to the same size of matrices (i.e., to the same values of $N_s$) accessible to the $T=0$ method, and the only limitation is given by the computational time, that grows in order to obtain accurate excited states. The actual increase of total time will depend on the temperature and on the size of the system, but our results for $N_s=8$, where the increase factor is around 3 for a range of temperatures that approaches the Mott transition endpoint, are quite promising. We believe anyway that the main use of this approach for multiorbital or cluster models can be to elucidate the really small temperature range, which is never easy to capture with other impurity solvers in the DMFT framework. This range is reasonably accessible with an affordable increase of computational time for the largest systems used in CDMFT and multiorbital DMFT.

We thank J. Tomczak, S.Biermann for discussions and for the QMC data shown in Fig.~3,
and N. Bl\"umer for kindly providing us with the data of Ref.~\cite{blumer}.
M.C. gratefully thanks the hospitality of \'Ecole Polytechnique, where this work has been mainly carried out and discussions with E. Koch. This work has been partially sponsored by CNR-INFM,
MIUR PRIN Prot. 200522492, CNRS-France and  \'Ecole Polytechnique.



\begin{thebibliography}{99}
\bibitem{revdmft} For a review, see e.g:
A.~Georges, G.~Kotliar, W.~Krauth, and M.~J.~Rozenberg, 
Rev. Mod. Phys. {\bf 68}, 13 (1996).

\bibitem{dmftmaterials} For reviews, see e.g:
K.~Held, I.~A.~Nekrasov, N.~Bl\"umer, V.~I.~Anisimov, and D.~Vollhardt, Int. J. Mod. Phys. B, {\bf 15}, 2611 (2001); A.~Georges, Lectures on the physics of
  highly correlated electron systems VIII (A.~Avella and F.~Mancini, eds.),
  American Institute of Physics Pub., 2004, (cond-mat/0403123);
G.~Kotliar, S.~Savrasov, K.~Haule, V.~Oudovenko, O.~Parcollet, and C.~Marianetti, 
Rev. Mod. Phys. {\bf 78}, 000865 (2006).

\bibitem{gk_dmft} A.~Georges and G.~Kotliar, Phys. Rev. B {\bf 45}, 6479 (1992).

\bibitem{ed} M.~Caffarel and W.~Krauth, Phys. Rev. Lett. \textbf{72}, 1545 (1994).

\bibitem{qmc} J.~E.~Hirsch and R.~M.~Fye, Phys. Rev. Lett. {\bf 56}, 2521 (1986).

\bibitem{ctqmc} A.~N.~Rubtsov, V.~V.~Savkin, and A.~I.~Lichtenstein, Phys. Rev. B  {\bf 72}, 035122 (2005); P.~Werner, A.~Comanac, L.~de' Medici, M.~Troyer, A.~J.~Millis, Phys. Rev. Lett. {\bf 97}, 076405 (2006).

\bibitem{nrg} R.~Bulla, Phys. Rev. Lett. {\bf 83}, 136 (1999).


\bibitem{nca} T.~Pruschke, M.~Jarrell and J.~Freericks,
{\sl Adv. Phys.} {\bf 42}, 187 (1995).

\bibitem{rotors} S.~Florens and A.~Georges, Phys. Rev. B {\bf 66},165111 (2002).

\bibitem{sef} M.~Potthoff, Eur. Phys. J. B {\bf 32}, 429 (2003).

\bibitem{prelovsek} J.~Jaklic and P.~Prelovsek, Phys. Rev. B {\bf 49}, 5065 (1994); J.~Jaklic and P.~Prelovsek, Adv. Phys. {\bf 49}, 1 (2000).

\bibitem{aichhorn} M.~Aichhorn, M.~Daghofer, H.~G.~Evertz, and W.~von der Linden 
Phys. Rev. B {\bf 67} 161103(R) (2003)

\bibitem{notageometrie} The non-interacting bath can be parametrized in equivalent 
alternative ways, as discussed in Ref. \cite{revdmft}

\bibitem{1d} M.~Capone, M.~Civelli, S.~S.~Kancharla, C.~Castellani and G.~Kotliar,
Phys. Rev. B {\bf 69}, 195105 (2004).

\bibitem{toschi} For a recent calculation of optical properties of the Hubbard model see A. Toschi, M. Capone, M. Ortolani, P. Calvani, S. Lupi, and C. Castellani, Phys. Rev. Lett. {\bf 95}, 097002 (2005).

\bibitem{notabethe} The linear proportionality between the hybridization 
function $\Delta(\iomn)$ (l.h.s. term of (\ref{selfcons})) and $G(\iomn)$ holds
 only for the Bethe lattice, but also in generic lattices the former quantity
is a given function of the latter, so that a more structured $G$ reflects in a similar property for $\Delta$.

\bibitem{golubvanloan} G.~H.~Golub and C.~F.~Van Loan, Matrix Computations,
John Hopkins University Press (Baltimore), 1993.

\bibitem{notalanczos}
$G^{(m)}$ is obtained as a continuous fraction with coefficients
determined from the Lanczos procedure
starting from the vector $f^{\dagger}_{\sigma}\vert m \rangle$ (or  $f_{\sigma}\vert m \rangle$).

\bibitem{jan} J.~M.~Tomczak, private communication

\bibitem{bulla}  R.~Bulla, T.~A.~Costi, D.~Vollhardt,
Phys. Rev. B {\bf 64}, 045103 (2001).

\bibitem{udo} J.~Joo and V.~Oudovenko,
Phys. Rev. B, {\bf 64}, 193102 (2001).

\bibitem{blumer} N.~Bl\"umer, Ph.D. Thesis, Universit\"at Augsburg, 2002.

\bibitem{cdmft} G.~Kotliar, S.~Y.~Savrasov, G.~Palsson and G.~Biroli,
Phys. Rev. Lett. {\bf 87}, 186401 (2001).

\bibitem{dca} T.~Mauer, M.~Jarrell, T.~Pruschke, and J.~Keller, 
Eur. Phys. J. B {\bf 13}, 613 (2000); M.~H.~Hettler, A.~N.~Tahvildar-Zadeh, 
M.~Jarrell, T.~Pruschke, and H.~R.~Krishnamurthy, Phys. Rev B {\bf 58}, 
R7475 (1998)

\bibitem{c60} M.~Capone, M.~Fabrizio, C.~Castellani, and E.~Tosatti, Science {\bf 296}, 2364 (2002);
Phys. Rev. Lett. {\bf 93}, 047001 (2004)


\bibitem{plaquette} M.~Civelli, M.~Capone, S.~S.~Kancharla, O.~Parcollet, and G.~Kotliar,
Phys. Rev. Lett. {\bf 95}, 106402 (2005); B.~Kyung, S.~S.~Kancharla, D.~S\'en\'echal, 
A.~-M.~S.~Tremblay, M.~Civelli, and G.~Kotliar, Phys. Rev. B {\bf 73}, 165114.



\end{thebibliography}
\end{document}